\documentclass[10pt,twocolumn,letterpaper]{article}

\usepackage{iccv}
\usepackage{times}
\usepackage{epsfig}
\usepackage{graphicx}
\usepackage{amsmath}
\usepackage{amssymb}

\usepackage{url}
\usepackage{amsmath,amssymb,graphicx}
\usepackage{mathtools,amssymb, bm}
\usepackage{booktabs, isotope}
\usepackage{lipsum} % Only used to generate random text.
\usepackage{times}
\usepackage{epsfig}
\usepackage{graphicx}
\usepackage{amsmath}
\usepackage{amssymb}
\usepackage{amsmath,graphicx}
\usepackage{amsmath,amssymb,amsfonts}
\usepackage[labelfont=bf,list=true,font=footnotesize, margin=0pt]{subcaption}
\usepackage[labelfont=bf, font=footnotesize, margin=0pt]{caption}
\usepackage{balance,cite}
\usepackage{algorithm}
\usepackage{algorithmic}
\usepackage{siunitx}
\usepackage{setspace,color, enumitem}
\usepackage{flushend}
\usepackage{balance}
\usepackage{sidecap}
\usepackage{float}
\usepackage{xr} 
\usepackage{animate}

\usepackage[switch]{lineno}

\usepackage{gensymb}

\usepackage[normalem]{ulem}
\newif\ifannotated

\annotatedtrue
\ifannotated
\newcommand{\delete}[1]{{\color{red}{\sout{#1}}}}

\newcommand{\margincomment}[1]{\marginpar{$\Rightarrow$\color{red}\fbox{\parbox{\linewidth}{\color{black}\scriptsize#1}}}}
\else
\newcommand{\delete}[1]{\ignorespaces}

\newcommand{\margincomment}[1]{}
\fi

\usepackage[pagebackref=true,breaklinks=true,letterpaper=true,colorlinks,bookmarks=false]{hyperref}

\iccvfinalcopy % *** Uncomment this line for the final submission

\ificcvfinal\fi

\newcommand{\mL}{\ensuremath{\mathbf L}}

\newcommand{\mY}{\ensuremath{\mathbf Y}}

\newcommand{\vl}{\ensuremath{\mathbf l}}

\newcommand{\vy}{\ensuremath{\mathbf y}}

\newcommand{\bphi}{\ensuremath{\bm{\varphi}}}
\newcommand{\mPhi}{\ensuremath{\bm{\Phi}}}

\begin{document}

%%%%%%%%% TITLE
% \title{FourierCam}
\title{A Simple Framework for 3D Lensless Imaging with Programmable Masks}

\author{Yucheng Zheng$^\dagger$, Yi Hua$^\ddag$, Aswin C. Sankaranarayanan$^\ddag$, M. Salman Asif$^\dagger$\\
University of California Riverside$^\dagger$, Carnegie Mellon University$^\ddag$ \\
{\tt\small yzhen069@ucr.edu,  huayi@cmu.edu,  saswin@andrew.cmu.edu, sasif@ece.ucr.edu}
}

\maketitle
\ificcvfinal\thispagestyle{empty}\fi

\begin{abstract}
Lensless cameras provide a framework to build thin imaging systems by replacing the lens in a conventional camera with an amplitude or phase mask near the sensor. Existing methods for lensless imaging can recover the  depth and intensity of the scene, but they require solving computationally-expensive inverse problems. Furthermore, existing methods struggle to recover dense scenes with large depth variations. In this paper, we propose a lensless imaging system that captures a small number of measurements using different  patterns on a programmable mask. In this context, we make three contributions. First, we  present a fast  recovery algorithm to recover textures on a fixed number of depth planes in the scene. Second, we consider the mask design problem, for programmable lensless cameras, and provide a design template for  optimizing the mask patterns with the goal of improving depth estimation. Third, we use a refinement network as a post-processing step to identify and remove artifacts in the reconstruction. These modifications are evaluated extensively with experimental results on a lensless camera prototype to showcase the performance benefits of the optimized masks and recovery algorithms over the state of the art. 
\end{abstract} 

%%%%%%%%% BODY TEXT
\section{Introduction}

% brief introduction to lensless cameras
%
Lensless cameras have traditionally been developed and used for imaging high-energy radiations such as X-ray and Gamma ray, where fabricating lenses  is either extremely expensive or physically unrealizable \cite{fenimore1978ura, busboom1998ura, cannon1980coded_aperture}. In the last few years, such lensless cameras have been proposed for imaging in visible and infrared wavebands as well; here, the absence of the lens provides advantages in thin form-factor imaging \cite{asif2017flatcam, boominathan2016lensless}, better depth-of-field tradeoffs in microscopy \cite{antipa2018diffusercam,adams2017rice_depth}, reduced cost in infrared/multi-spectral imaging \cite{sony2020infrared,monakhova2020spectral}, and provides the ability to do inference from coded measurements \cite{sony2020infrared,tan2019face}. 

% depth recovery in lensless - principles
Depth estimation is an integral part of lensless imaging.
The basic principle of a lensless camera is that the sensor records the summation of the measurements associated with the scene points. Each scene point casts its own unique sensor measurement, depending on its three-dimensional (3D) spatial location and radiance. Thus, the 3D scene information is encoded in the sensor measurements, but its recovery requires us to solve a nonlinear inverse problem. Further, it is important to model this depth dependence in the measurements especially since ignoring it results in significant reduction in the quality of reconstructions.
% depth dependence invariably results in reconstructions that are of a poor quality.
%, even in simple scenarios like a planar scene scene consists of a single plane, we need precise depth information of the plane to recover the image intensity. 

\if 0 
\begin{figure}[t]
	% \footnotesize
	% \tiny
	\scriptsize
	\centering
	\renewcommand{\tabcolsep}{1.0pt} % adjust horizontal space
	\renewcommand{\arraystretch}{1.0} % adjust vertical space
	\begin{tabular}{cc}
	SweepCam \cite{hua2020sweep} & Ours \\
  \begin{frame}{}
    \animategraphics[loop,autoplay,width=0.45\linewidth,nomouse]{0.8}{figures/sweepCam_scene_net_depth_}{00}{08}
  \end{frame}
    &
  \begin{frame}{}
    \animategraphics[loop,autoplay,width=0.45\linewidth,nomouse]{0.8}{figures/blockDiagonal_scene_net_depth_}{00}{08}
  \end{frame} \\
  \begin{frame}{}
    \animategraphics[loop,autoplay,width=0.45\linewidth,nomouse]{0.8}{figures/sweepCam_scene_toys_depth_}{00}{08}
  \end{frame}
 &
  \begin{frame}{}
    \animategraphics[loop,autoplay,width=0.45\linewidth,nomouse]{0.8}{figures/blockDiagonal_scene_toys_depth_}{00}{08}
  \end{frame}   
  
\end{tabular}
	\label{fig:preface} %% label for entire figure
%  	\vspace{-10pt}
\caption{Examples of 3D images reconstructed  from eight sensor measurements using SweepCam~\cite{hua2020sweep} and our proposed method. \textit{This is a video figure, best viewed in Adobe$^\text{\textregistered}$ Reader.}}\label{fig:teaser}
\end{figure}
\fi

\newcommand{\figwidth}{0.23\linewidth}
\begin{figure}[t]
    \centering
    \scriptsize
	\setlength\tabcolsep{1pt}
	\begin{tabular}{cc|cc}
		27.3mm &
		% 42.9mm &
		100.0mm &
		91.3mm &
		% 42.9mm &
		311.1mm \\
		\rotatebox{90}{\parbox{1cm}{\centering  SweepCam}}
		\includegraphics[width=\figwidth,keepaspectratio]{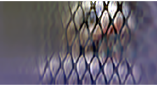} &
		\includegraphics[width=\figwidth,keepaspectratio]{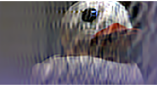}&
		\includegraphics[width=\figwidth,keepaspectratio]{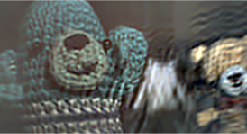} &
		\includegraphics[width=\figwidth,keepaspectratio]{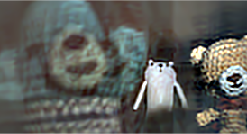}
		\\
		\rotatebox{90}{\parbox{1cm}{\centering  Ours}}
		\includegraphics[width=\figwidth,keepaspectratio]{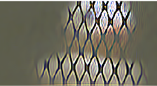} &
		\includegraphics[width=\figwidth,keepaspectratio]{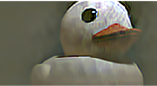}&
		\includegraphics[width=\figwidth,keepaspectratio]{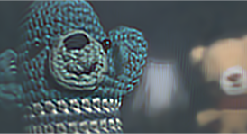} &
		\includegraphics[width=\figwidth,keepaspectratio]{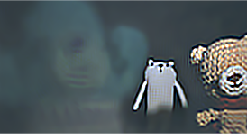}
	\end{tabular}
	\caption{Examples of two 3D scenes reconstructed at different depth planes from eight sensor measurements using SweepCam~\cite{hua2020sweep} and our proposed method. }
	\label{fig:teaser}
\end{figure}

% existing lensless depth estimation
Existing methods for 3D lensless imaging can be divided into two categories. One category of methods estimate the 3D scene with a single image measurement \cite{asif2017lensless3D, zheng2019depth , antipa2018diffusercam,boominathan2020phlatcam,adams2017rice_depth}. These methods jointly estimate the image and depth map of a 3D scene by solving an optimization problem using iterative techniques. 
Since the number of variables to be estimated is much larger than the number of measurements, the recovery problem is severely under-determined and the methods rely on some prior knowledge about the 3D scene. For instance, \cite{antipa2018diffusercam} assumes that 3D volume is sparse and solves an $\ell_1$ norm-based optimization problem to estimate a 3D image. Another drawback of these optimization-based methods is their large computational cost and run time. The second category of methods capture multiple measurements, each with a different mask, which makes the 3D recovery of dense scenes possible \cite{hua2020sweep,yamaguchi2019lensless}. 
The most related work that concerns designing mask patterns for multiple measurements is \textit{SweepCam}\cite{hua2020sweep}, which captures multiple measurements of the same scene using a programmable, shifting mask and estimates one plane in the 3D scene at a time. The recovery of a single plane is much faster than joint recovery of the entire 3D scene, but the number of mask patterns needed to achieve artifact-free reconstruction of a single depth plane can be large (in the range of 100--400). Our proposed method falls in the second category, but instead of estimating a single plane, we recover the 3D scene at multiple depth planes jointly from a smaller number of measurements, using a simple algorithm. A comparison of our method and SweepCam is shown in Figure~\ref{fig:teaser} for the recovery of two 3D scenes using eight sensor measurements. 

\begin{figure*}[t]
	\centering
	\includegraphics[width=1\linewidth, keepaspectratio]{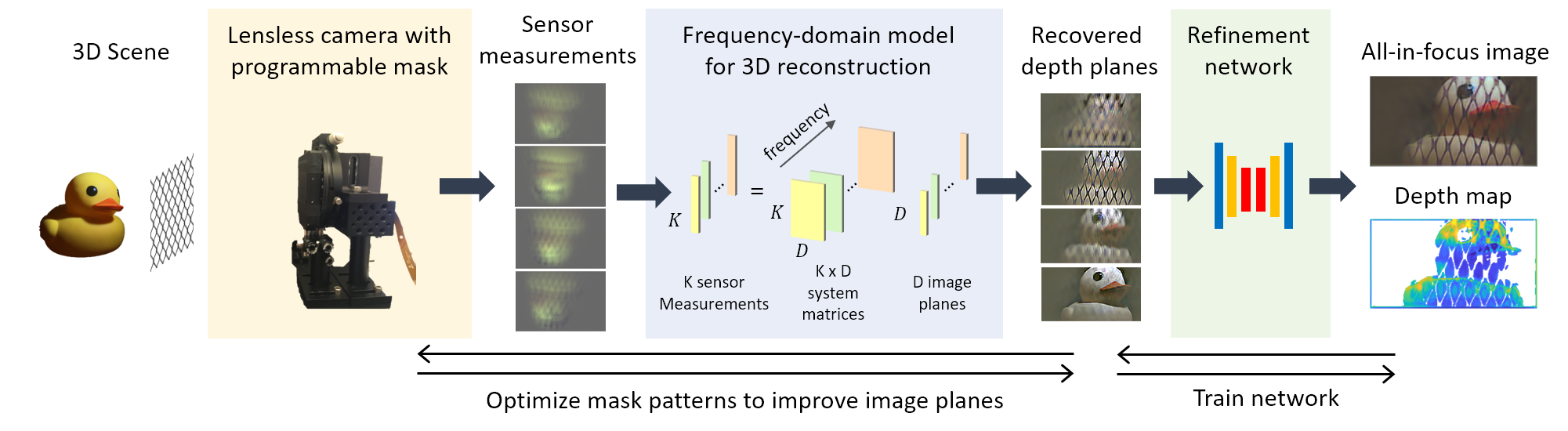}

	\caption{An overview of the proposed method. The lensless camera captures multiple measurements with a programmable mask. We reconstruct multiple image planes in the 3D scene by solving multiple small systems of equations in parallel (one for each frequency component). Using the recovery algorithm as a differentiable function, we learn the mask patterns to improve the estimates of image planes. We further refine the estimated image planes and convert them into an all-in-focus image and a depth map using a trained refinement network. }
	\label{fig:overview}
\end{figure*}

% overview of our method 
%We present a simple framework to recover 3D scenes from lensless measurements captured with a programmable mask. We represent 3D scenes using multi-plane images and sensor measurements with a given mask pattern as the summation of measurements of all the depth planes. 
The proposed algorithm exploits the well-known fact that convolutional systems are diagonalized in the Fourier domain. Given that the measurement operator associated with each scene depth is well-approximated as a convolution, we show that the joint depth and image recovery problem can be significantly simplified when we model the scene intensities in the spatial frequency domain. Hence, instead of solving a single large linear system for image intensity across all of the depth planes, we solve several small linear systems in parallel, with each system associated with a different frequency coefficient.

To further improve the quality of estimated image planes, we implement the forward imaging operation and the recovery algorithm as a differentiable network and optimize over the mask patterns to minimize the reconstruction loss. 
We validate in real experiments that the learned mask patterns provide significantly better estimates of image planes for a variety of scenes compared to other commonly used mask patterns. Finally, we train a neural network to remove any artifacts in the estimated planes and convert them into an all-in-focus image and a continuous-valued depth map.
%

%Finally, we built a prototype camera to capture lensless measurements with a programmable mask for a variety of 3D scenes. Our results demonstrate that our proposed method provides robust and accurate recovery of intensity and depth for natural and dense 3D scenes.

% \vspace{-10pt}
% contributions
\paragraph*{Contributions.}
The main contributions of this paper are as follows. 
\begin{itemize}[leftmargin=*,noitemsep,topsep=0pt]
	\item We present a fast algorithm to recover multi-plane images for the 3D scene from lensless measurements in the Fourier domain. 
	\item We implement a multi-plane lensless camera and Fourier-domain recovery algorithm as a differentiable network to optimize mask patterns for the lensless camera. 
	\item We train a neural network to map the estimated multi-plane images into an all-in-focus image and a continuous-valued depth map.
	\item We built a prototype lensless camera with a programmable mask and report several experiments to validate our proposed methods and a comparison with existing methods. 
\end{itemize}

\paragraph*{Limitations.} Despite the advantages listed above, the mask-based lensless camera that we propose here has some limitations. One main limitation, common to many lensless cameras, is the light throughput and limited dynamic range of the sensor. 
A second set of limitations stem from the assumption of the convolutional model, a common assumption in 3D lensless imaging work.
The convolutional model fails to account for small sensor area that crop the measurements, non-Lambertian scenes, and the sensor's angular response. 
In our experiments, we observe some artifacts in those scenarios, but they are usually local; for example, artifacts from sensor cropping only affect image boundaries and specular reflections only produce artifacts in local regions, and do not affect other parts of images.
Finally, we assume the scene is static in the duration that multiple measurements are captured. 
However, this work significantly reduces the number of measurement required, and hence the duration in which the scene is required to be static, compared to previous methods \cite{hua2020sweep}.

\section{Related Work} 
% introduction on FlatCam and diffusercam
A pinhole camera is a classical example of lensless imaging. 
FlatCam \cite{asif2017flatcam} is an extended version of a pinhole camera that uses a mask that has multiple pinholes. 
The sensor captures linear measurements of the scene and the system requires careful calibration and computational algorithm to reconstruct the target scene. 
The lensless imaging system underlying FlatCam has been used in face detection \cite{tan2019face}, privacy protection \cite{canh2019iccv} and fluorescence microscopy \cite{adams2017rice_depth}. Recently, deep learning-based recovery methods have also been introduced for FlatCam in \cite{khan2019iccv}.

In contrast to FlatCam, where an amplitude mask is used, DiffuserCam \cite{antipa2018diffusercam} places a phase mask on top of the sensor. The imaging system exploits the shift-invariant property of the PSF and models the measurements as the convolution between the scene and corresponding PSF. 
DiffuerCam has also been used to recover other types of high-dimensional signals, such as video \cite{antipa2019still} and hyperspectral images \cite{monakhova2020spectral}.
Methods for learning-based approaches have been introduced in \cite{monakhova2019learning}.  

% single mask: 1. joint image and depth reconstruction (greedy + diffusercam). 2. 3D volume recon
Algorithms for depth estimation or 3D image reconstruction with a single mask-based FlatCam and DiffuserCam can be divided into two categories. One approach solves a sparse recovery problem based on a 3D grid \cite{adams2017rice_depth, antipa2018diffusercam}. These methods are often used to recover sparse images in fluorescence microscopy where all the light point sources in the field of view can be detected by the sensor. The main drawback of these iterative 3D reconstruction algorithms is that they often require large time and space complexity for recovery. The other approach jointly estimates the image intensity and depth of every point/angle in the image \cite{asif2017lensless3D,zheng2019depth}. These methods assume angular sampling of 3D scenes in which only one light source exists at every angle and a greedy or nonlinear optimization algorithm to estimate the intensity and the depth map of the 3D scene.  

% multi masks (sweepcam). main disadvantage: artifacts...not reconstructing a clear 3D volume
SweepCam \cite{hua2020sweep} is an extension to FlatCam that reconstructs image texture and depth map using multiple sensor measurements captured by shifting the mask pattern. The shifting mask pattern provides a depth-dependent disparity of different planes in the 3D scene for every measurement instance, which makes it possible to recover a focused image for a specific depth plane. SweepCam reconstructs one depth at a time and generates depth map using local contrast information of reconstructed depth planes. The main disadvantage of SweepCam is that it requires a very large number of sensor measurements, often in hundreds, to recover the depth planes without artifacts. In this paper, we use a similar imaging setup as SweepCam \cite{hua2020sweep} but use only eight to ten sensor measurements to recover the depth planes jointly. 

Several methods for joint design of optics and reconstruction algorithms have been recently proposed in \cite{chang2019deep, wu2019phasecam, boominathan2020phlatcam, sun2020learning, metzler2020hdr, gordon2020deep , sitzmann2018deep}. The main design principle is to represent sensor measurements as a differentiable function of the scene image and reconstruction of image as a differentiable neural network, after which parameters for the optical elements and the neural network can be learned jointly in an end-to-end manner. In our proposed framework, we recover 3D scene as a regularized least squares problem that has a closed-form solution. The main motivation for using a simple solver instead of training a deep network for reconstruction is to build a well-conditioned system that is generalizable to arbitrary scenes. We then train a separate refinement network that maps the estimated multi-plane images to an all-in-focus image and a continuous-valued depth map.

\section{Methods}

\subsection{Multi-Plane Lensless Imaging Model} 
\label{sec:general_imaging_model}

% describe your scene and sensor model
%
We introduce the model underlying our lensless imager, where 3D scenes are  represented using multi-plane images.
Under some mild conditions on the scene and sensor sizes, we can represent sensor measurements as a summation of every scene plane convolved with its respective depth-dependent point spread function (PSF). The PSF depends on the mask pattern, which we can change to capture multiple sensor measurements and recover the 3D scene. 

Let us assume that our lensless camera consists of a programmable amplitude mask placed at a distance $d$ from a planar sensor array. Let us further assume that the sensor is placed at the origin of a Cartesian coordinate system indexed by $(u,v,z)$, where $(u,v)$ denotes the horizontal and vertical coordinates and $z$ denotes the depth. 
In a mask-based lensless camera, every sensor pixel records a linear combination of light coming from all points in the 3D scene. Let us assume that light rays originating from scene point $(u,v,z)$ have same effective brightness $l(u,v,z)$ with respect to every sensor pixel.  
Suppose we use an amplitude mask whose attenuation at $(u,v)$ in the mask plane is represented as $\varphi(u,v)$.
Under these assumptions, the sensor measurement recorded at pixel $(u,v)$ is given as 
\begin{equation}
y(u,v) = \int_z \left[\int_{u_0, v_0} l(u_0, v_0, z) \varphi\left( \tilde u, \tilde v\right) du_0 dv_0 \right]~dz,
\label{eq:integration}
\end{equation}
where $\tilde u = \frac{z-d}{z} u + \frac{d}{z} u_0$, $\tilde v = \frac{z-d}{z} v+ \frac{d}{z}  v_0$. 
A simple derivation shows that the inner integration in \eqref{eq:integration} for a fixed $z$ can be represented as a convolution of two functions \cite{hua2020sweep,antipa2018diffusercam}:  
\begin{align}
y(u,v) & = \int_z \left[\int_{{u_0,v_0}} \kern-10pt l_z( u_0, v_0) \varphi_z(u-u_0, v-v_0) d u_0 d v_0 \right]dz \notag \\
& =  \int_z [l_z \ast  \varphi_z]( u, v)  dz, \label{eq:conovlution_tilde}
\end{align}
where $l_z(u,v)=l(\frac{d-z}{d}u,\frac{d-z}{d}v)$ denotes the scene plane at depth $z$ and $\varphi_z(u,v) = \varphi( \frac{z-d}{z} u, \frac{z-d}{z} v)$ denotes the PSF for any point at depth $z$.

% discrete sampling of 3D space 
Let us assume that the sensor array has $M$ pixels on a uniform grid. By discretizing the system in \eqref{eq:conovlution_tilde} along $(u,v,z)$, we can write the sensor measurements as a summation of 2D linear convolutions: 
\begin{equation}\label{eq:convolution_vector}
    \vy = \sum_z \bphi_z * \vl_z +\mathbf{e}, %  \equiv \sum_z \mathcal{T}_{\varphi_z}  \vl_z,
\end{equation}
where $\vy$ denotes an array with $M$ sensor measurements $y(u,v)$, $\vl_z$ denotes image-plane intensity values $l_z(u,v)$, $\bphi_z$ denotes the PSF for depth $z$, and $\mathbf{e}$ denotes the sensor noise. 
For the ease of exposition, we ignore the additive noise term in the equations below, but we include noise in our experiments. 
If the sensor size is large enough so that the response of every light source in the scene is completely recorded by the sensor, then we can treat the linear convolution in \eqref{eq:convolution_vector} as circular convolution. We can diagonalize the system in \eqref{eq:convolution_vector}, using the convolution theorem, as
\begin{equation}
\mathcal{F}(\vy) = \mathcal{F}(\bphi_z) \odot \mathcal{F}(\vl_z) \Rightarrow  \mY =   \sum_z  \mPhi_z  \odot  \mL_z,
\label{eq:single_plane_convf}
\end{equation}
where $\mathcal{F}(\cdot)$ denotes the 2D Fourier transform operator, $\mY$, $\mPhi_z$, and $\mL_z$ denote 2D Fourier transforms of $\vy$, $\bphi_z$ and $\vl_z$, respectively, and $\odot$ denotes an element-wise multiplication operator. All the arrays in \eqref{eq:single_plane_convf} have $M$ entries.

Since we use a programmable mask, we can capture multiple measurements of the scene using different mask patterns. Let us represent the sensor measurements captured using $k$th mask pattern as
\begin{equation}
    \vy^k = \sum_z \bphi_z^k * \vl_z ~~\overset{\mathcal{F}}{\longrightarrow}~~ \mY^k = \sum_z \mPhi_z^k \odot \mL_z. \label{eq:progMaskMeasurements}
\end{equation}

\subsection{Fast Algorithm for Multi-Plane Reconstruction}
\label{sec:reconstruction_algorithm}

In this section, we discuss a Fourier-domain algorithm to recover 3D scene using multiple lensless measurements in \eqref{eq:progMaskMeasurements}. This method is an adaptation of the classical frequency-domain multi-channel deconvolution method \cite{galatsanos1991least} to the lensless imaging setup. Let us assume that we are given $K$ sensor measurements as $\vy^k$ for $k = 1,\ldots, K$. Let us further assume that the 3D scene consists of $D$ planes represented as  $\vl_{1},\ldots, \vl_D$. To simplify the notation, we will use $\bphi_1, \ldots, \bphi_D$ to denote PSFs for different depths. 
To recover the image planes in the Fourier domain, we can solve the following regularized least-squares problem:
\begin{equation}
\underset{\mL_1,\ldots,\mL_D}{\text{min}}  \; \sum_k \|\mY_k - \sum_z \mPhi_z \odot \mL_z \|_F^2 +  \tau \sum_z \|\mL_z\|_F^2,  \label{eq:LS_full}
\end{equation}
which involves solving for $MD$ unknowns using $MK$ measurements and can be computationally expensive to solve for large values of $M$, $K$, and $D$. 

Fortunately, we can simplify the recovery problem by separating the optimization problem in \eqref{eq:LS_full} into $M$ independent problems, each defined over a single frequency coefficient in all the depth planes. Note that because of the diagonal structure in the Fourier domain, we can separate the measurements for a frequency coefficient $\omega_{m}$ as 
\begin{equation}\label{eq:separatedSystemOmega}
\footnotesize 
        \begin{bmatrix}
        \mY^1(\omega_m)  \\
        \vdots  \\
        \mY^K(\omega_m) 
        \end{bmatrix} 
        =
        \begin{bmatrix}
        \mPhi_1^1(\omega_m)  & \ldots &  \mPhi_D^1(\omega_m)  \\
        \vdots & \ddots & \vdots \\
        \mPhi_1^K(\omega_m)  & \ldots & \mPhi_D^K(\omega_m) 
        \end{bmatrix}
        \begin{bmatrix}
        \mL_1(\omega_m)  \\
        \vdots \\
        \mL_D(\omega_m)
        \end{bmatrix}.
\end{equation}
We can rewrite this system in a compact form as 
\begin{equation}
    \mY_{\omega_m} = \mPhi_{\omega_m} \mL_{\omega_m}, \label{eq:separatedSystem}
\end{equation}
$\mY_{\omega_m}$ is a complex vector of length $K$, $ \mPhi_{\omega_m}$ is a $K\times D$ complex matrix, and $\mL_{\omega_m}$ is the unknown vector of length $D$. 
The original optimization problem can be written as a summation of $M$ independent optimization problems as
\begin{equation}
    \underset{\mL_{\omega_1},\ldots,\mL_{\omega_M}}{\text{min}}  \; \sum_{\omega_m} \|\mY_{\omega_m} - \mPhi_{\omega_m} \mL_{\omega_m} \|_F^2 +  \tau \|\mL_{\omega_m}\|_F^2.   \label{eq:LS_parallel}
\end{equation}
The solution for any $\omega_m$ can be written in a closed-form as 
\begin{equation}
    \widetilde{\mL}_{\omega_m} = ( \mPhi_{\omega_m}^*\mPhi_{\omega_m} + \tau I)^{-1} \mPhi_{\omega_m}^* \mY_{\omega_m}, \label{eq:LS_parallel_sol} 
\end{equation}
which we can compute by either directly inverting the $D\times D$ matrices or using an iterative method such as conjugate gradients \cite{golub1996matrix}. Since the computations for all the $\omega_m$ are independent of one another, we can solve \eqref{eq:LS_parallel_sol} in parallel, which provides a fast recovery algorithm. To recover the 3D image planes, we can apply inverse Fourier transform on reconstructed frequency-domain depth planes. In practice, we adjust $\tau$ for each $\omega_m$ according to the Frobenius norm of the corresponding system matrix. Our method recovers multiple image planes from multiple measurements using a closed-form solution. In fact, if $K=D=1$, then our method is equivalent to the standard Wiener deconvolution. 

\paragraph*{Practical considerations.}
To ensure stable recovery of \textit{arbitrary} 3D scenes, we desire  the matrices in \eqref{eq:LS_parallel_sol} to be invertible for all frequencies $\omega_m$. In practice, we can resolve this issue in two ways. 
First, the energy of the Fourier coefficients for natural images is mostly concentrated at a small number of frequencies; therefore, we can recover the image planes reliably as long as the matrices corresponding to the significant frequencies are well-conditioned. In Sec.~\ref{sec:mask_optimization}, we discuss an approach to improve the conditioning of the system and the estimation of the image planes by optimizing the mask patterns. 
Second, even though the invertibility condition requires the number of depth planes $(D)$ to be at most the number of mask patterns $(K)$, we have the flexibility to choose which depth planes  to recover. We can adjust the location of depth planes according to the scene or select the depth planes that provide the best recovery performance. In our experiments, we observed that sampling depth planes uniformly in $\alpha = 1-\frac{d}{z}$ parameter provides best reconstruction, where $\alpha \in [0, 1]$ maps to $z \in [d,\infty]$.

\paragraph*{Relation to existing methods.}
In our proposed method, we use multiple mask patterns to recover multiple depth planes in a 3D scene. Recovery of a 3D scene using a single mask pattern ($K=1$) is possible, but it remains a challenging problem. Existing methods for 3D lensless imaging from a single sensor image either assume sparse prior on the 3D scene and solve an $\ell_1$-regularized problem over the 3D volume \cite{antipa2018diffusercam,boominathan2016lensless} or solve a nonlinear inverse problem to jointly recover the intensity and depth of scene \cite{zheng2019depth,asif2017lensless3D}. Both of which are computationally expensive. SweepCam \cite{hua2020sweep} recover a single depth plane using a ``focusing'' operation. We can show that the focusing operation for a shifting mask pattern in SweepCam is equivalent to solving the system in \eqref{eq:separatedSystemOmega} for one depth at a time in the frequency domain. Mathematically, it is equivalent to estimating a single frequency $\omega_m$ for plane at depth $z$, which can be written as the following scalar equation: 
\begin{equation}
\widetilde{\mL}_z(\omega_m) = \frac{\sum_k \mPhi^k_z(\omega_m)^* \mY^k(\omega_m)}{( \sum_k \mPhi^k_z(\omega_m)^*\mPhi_z^k(\omega_m) + \tau)}. \label{eq:sweepcam} 
\end{equation} 
In the experiment section, we present a detailed comparison between the performance of our proposed method, depth pursuit in \cite{asif2017lensless3D}, and SweepCam \cite{hua2020sweep}; our results demonstrate that our method outperforms existing methods.

\subsection{Learning Mask Patterns} \label{sec:mask_optimization}

To improve the quality of estimated image planes, we optimize the mask patterns by implementing the multi-plane image recovery algorithm as a differentiable network. 
We build a computation graph that implements the  imaging model in \eqref{eq:progMaskMeasurements} and the fast recovery algorithm, as illustrated in Figure~\ref{fig:overview}. 
The optimization variable is a $K\times P_u\times P_v$ tensor that has $K$ mask patterns each of size  $P_u\times P_v$ (we used $P_u=P_v = 63$ in our experiments). We minimize the mean squared error (MSE) between the input and reconstructed image planes with respect to the mask patterns via backpropagation. 
We use 50 scenes from NYU~\cite{silberman2012NYU_depth} dataset as training data to optimize the masks. 

To implement the  imaging model in \eqref{eq:progMaskMeasurements}, we first perform linear interpolation to compute the PSF of every mask for $D$ predefined depth planes that are uniformly sampled along $\alpha = 1-\frac{d}{z}$. This provides us a $K\times D\times M_u\times M_v$ tensor, where $M_u\times M_v$ is the size of the sensor (i.e., $M = M_uM_v$). 
Then we generate sensor measurements for given training image planes using the convolution model in \eqref{eq:progMaskMeasurements}. 
We add independent instances of Gaussian noise in the sensor measurements during mask optimization. 
The reconstruction operator provides $D$ estimated image planes from the simulated measurements by solving the problem in \eqref{eq:LS_parallel}. As explained in Sec~\ref{sec:reconstruction_algorithm}, we can solve independent least-squares problems for all the frequencies and then reconstruct the image planes with an inverse Fourier transform. We optimize the masks for 300 epochs using Adam optimizer \cite{kingma2014adam} with the learning rate of 0.01.

We use a liquid crystal on Silicon (LCoS) spatial light modulator (SLM) as a programmable mask and restrict the mask patterns to be binary during training. The phase retardation in LCoS has a strong spectral dependence, which makes it hard to implement a desired continuous-valued pattern consistently across a span of wavelengths. This is not an issue for binary patterns since we can saturate phase retardation across all wavelengths.
We represent the mask as a zero-centered sigmoid function of a continuous-valued optimization variable, which keeps the mask values in the range [-1,1] during optimization. We increase the slope of the sigmoid function at every epoch, which pushes the mask values closer to -1 or 1, and we finally set them to -1/1 at the end of optimization. 

\subsection{Refinement Network and Post-Processing}
\label{sec:denoiser_description}
To further enhance the image quality and depth accuracy of the estimated planes, we train a neural network using the U-Net architecture \cite{olaf2015unet} that maps the estimated multi-plane images to an all-in-focus image and a continuous-valued depth map. The U-Net accepts all the color channel of the multi-plane image stack and generates an RGBD tensor \cite{wu2019phasecam,chang2019deep}.  
To train the U-Net parameters, we generated synthetic multi-plane images using NYU depth dataset \cite{silberman2012NYU_depth}. We first scale the depth of scenes in the NYU dataset linearly in the 35mm to 200mm range and then quantize the scenes to create 100 depth planes. Then we simulate the sensor measurements using the imaging model in \eqref{eq:progMaskMeasurements} and reconstruct images at $D=8$ depth planes by solving the problem in \eqref{eq:LS_parallel}. The reconstructed image planes are fed into the U-Net as a single tensor with $D$ RGB planes, and U-Net provides an output RGBD tensor. The U-Net loss function is defined as the mean squared error between the ground-truth and the output RGBD tensors. We used 380 scenes to train the network for 200 epochs.
In our experiments, we compare the U-Net results against a model-based approach in which we first denoise the  estimated image at every depth using BM3D \cite{dabov2007bm3d}, and then assign each pixel the depth value that provides maximum local contrast in the chosen depth plane.

\section{Simulation Results}

\renewcommand{\figwidth}{0.18\linewidth}

% - Programmable Masks (Mask patterns: MLS masks and random binary masks)
\begin{figure}[!t]
	\begin{subfigure}{1\linewidth}
	\setlength\tabcolsep{1pt}
	\scriptsize 
	\centering
	\begin{tabular}{c|cccc}
		% 		original image &
		Original &
		MLS &
		Random &
		Shifted MLS & 
		Learned\\
		\rotatebox{90}{\parbox{1.5cm}{\centering  image}}
		\includegraphics[width=\figwidth,keepaspectratio]{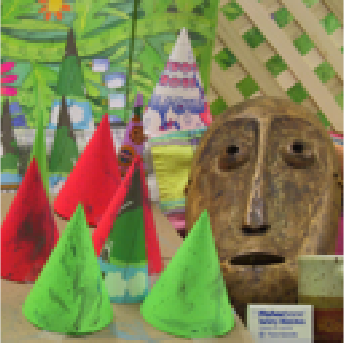} &
		\includegraphics[width=\figwidth,keepaspectratio]{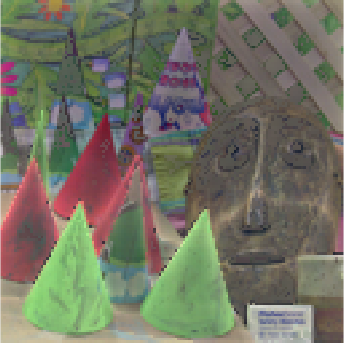} &
		\includegraphics[width=\figwidth,keepaspectratio]{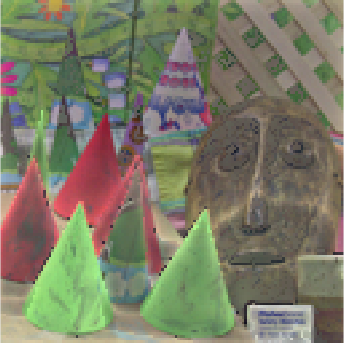} &
		\includegraphics[width=\figwidth,keepaspectratio]{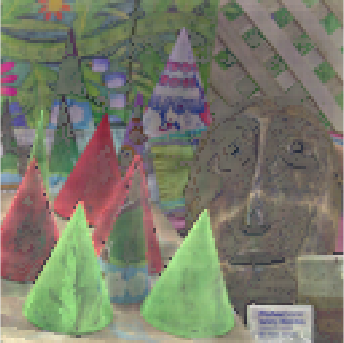}
		& 
		\includegraphics[width=\figwidth,keepaspectratio]{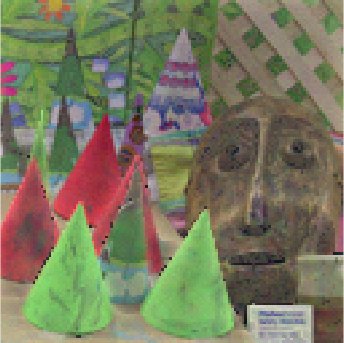}
		\\
		\rotatebox{90}{\parbox{1.5cm}{\centering depth}}
		\includegraphics[width=\figwidth,keepaspectratio]{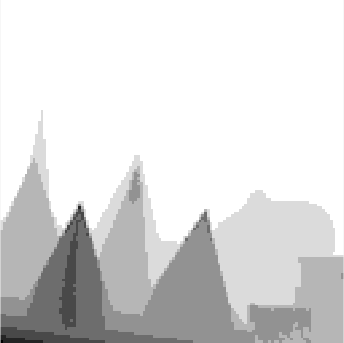} &
		\includegraphics[width=\figwidth,keepaspectratio]{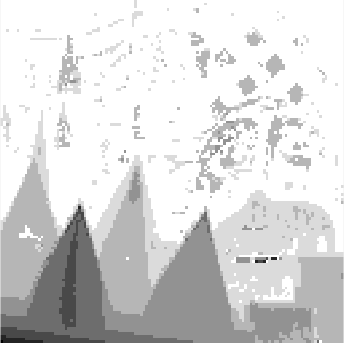} &
		\includegraphics[width=\figwidth,keepaspectratio]{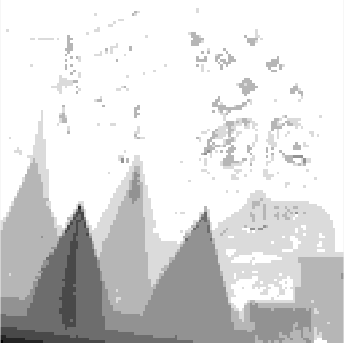} &
		\includegraphics[width=\figwidth,keepaspectratio]{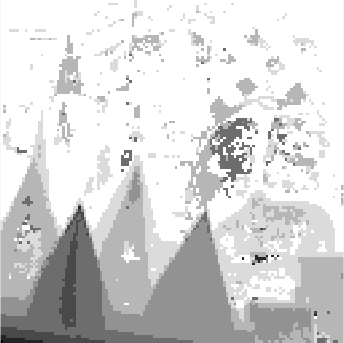} 
		& 
		\includegraphics[width=\figwidth,keepaspectratio]{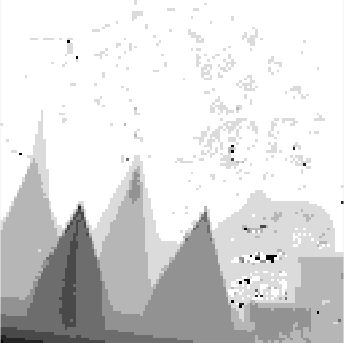}		
		\\
	\end{tabular}
	\caption{Reconstructed all-in-focus images and depth maps for cones with $K=8$.}
	\end{subfigure}
	\vspace{15pt} 
	
	\begin{subfigure}{0.49\linewidth}
		\includegraphics[width=1\linewidth,keepaspectratio]{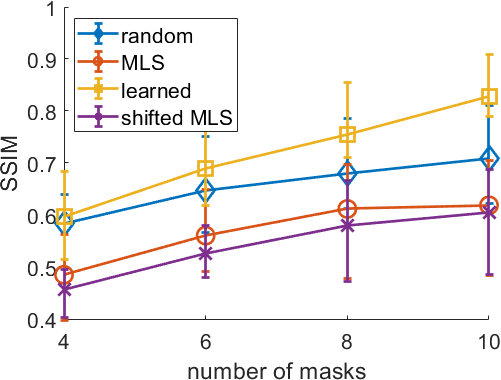}
		\caption{SSIM for all-in-focus image}
	\end{subfigure}
	\begin{subfigure}{0.49\linewidth}
		\includegraphics[width=1\linewidth,keepaspectratio]{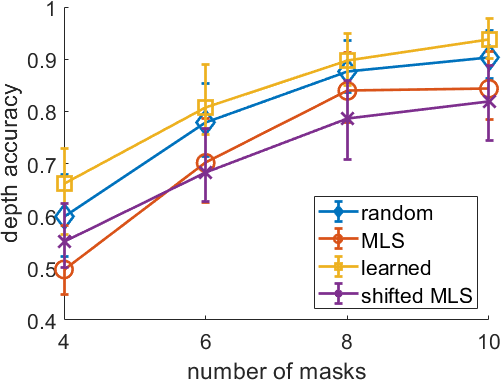}
		\caption{Accuracy for estimated depth}
	\end{subfigure}
    \vspace{10pt}
	\caption{Comparison of different types and number of masks. (a) Reconstructed all-in-focus images and depth maps for cones with $K=8$ measurements. (b,c) Average SSIM of recovered all-in-focus images and accuracy of estimated depth for five test scenes. Quality of reconstruction improves as the number of masks increases, and learned mask patterns outperform other mask patterns. }
	\label{fig:numMask_plot}
\end{figure}

\paragraph*{Simulation setup.}
To validate the proposed algorithm and learn the mask patterns, we simulated an imaging system for the camera prototype discussed in Sec.~\ref{sec:experiments}. We used a $256\times256$ sensor array with a mask placed 10.51mm away. 
The number of features in each mask pattern is $63 \times 63$ and each feature size is $36\mu$m. 
We evaluate the algorithm on five scenes selected from Middlebury dataset (cones, books, piano, playroom, and playtable) \cite{scharstein2001stereo}. 
The spatial resolution of each image is $128\times128$. We rescale the depth values in every scene to 35--380mm range and quantize the resulting depth into eight planes. We add Gaussian noise at 40dB SNR to the sensor measurements. 
After we reconstruct the image planes of the 3D scene using the proposed algorithm, we generate an all-in-focus image and depth map using local contrast in which we pick the depth with the largest local contrast among all planes for every pixel. We evaluate the quality of recovered all-in-focus images using structural similarity index (SSIM). We evaluate the quality of the estimated depth map using depth accuracy that we define as the ratio of pixels that are assigned the correct depth to those assigned incorrect depth.

\paragraph*{Number and types of mask patterns.}
We evaluate the performance of our proposed method for different types and number of mask patterns $(K)$. We test four different types of mask patterns: random binary masks, separable MLS  masks \cite{golomb2017shift}, shifted MLS masks as used in SweepCam \cite{hua2020sweep}, and learned mask patterns that are optimized according to the method discussed in Sec.~\ref{sec:mask_optimization}. 
We test $K=\{4,6,8,10\}$ for each mask pattern with five scenes, each of which have $D=8$ depth planes. Our test includes both the cases for under-determined $(K<D)$ and over-determined $(K>D)$ systems. 
We use sensor measurements with eight masks to reconstruct the multi-plane image stacks as described in \eqref{eq:LS_parallel_sol}. Then we convert image planes to an all-in-focus image and a depth map using local contrast. We present examples of reconstructed images and average performance curves over five test scenes in terms of SSIM and depth accuracy in Figure~\ref{fig:numMask_plot}. The reconstruction quality of image and depth improves as $K$ increases. We observe that the learned masks provide significantly better reconstruction for intensity and depth estimates compared to MLS, random, and shifted MLS. 
Incorrect depth estimates cause model mismatch, which in turn cause artifacts in the reconstructed intensity images that can be seen in Figure~\ref{fig:numMask_plot}(a). 
Additional examples of reconstructed all-in-focus images and depth maps are  in the supplementary material.

\section{Experiments with Camera Prototype}\label{sec:experiments}

\begin{figure*}[t]
	\setlength\tabcolsep{1pt}
	\centering
    \footnotesize
	\begin{tabular}{c|ccccc}
			camera view &
		near plane &
		middle plane &
		far plane &
		all-in-focus image & 
		depth map
		\\
		\includegraphics[width=0.122\linewidth,keepaspectratio]{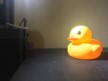} &
		\includegraphics[width=0.168\linewidth,keepaspectratio]{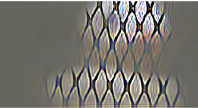} &
		\includegraphics[width=0.168\linewidth,keepaspectratio]{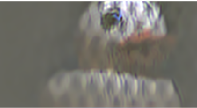} &
		\includegraphics[width=0.168\linewidth,keepaspectratio]{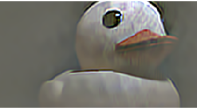} &
		\includegraphics[width=0.168\linewidth,keepaspectratio]{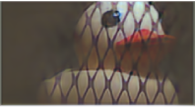} &
		\includegraphics[width=0.205\linewidth,keepaspectratio]{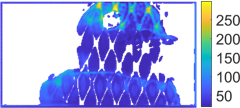}
\\
		\includegraphics[width=0.122\linewidth,keepaspectratio]{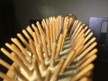} &
		\includegraphics[width=0.168\linewidth,keepaspectratio]{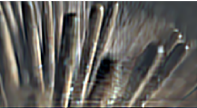} &
		\includegraphics[width=0.168\linewidth,keepaspectratio]{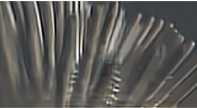} &
		\includegraphics[width=0.168\linewidth,keepaspectratio]{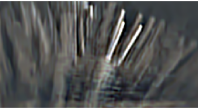} & 
		\includegraphics[width=0.168\linewidth,keepaspectratio]{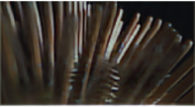} &
		\includegraphics[width=0.205\linewidth,keepaspectratio]{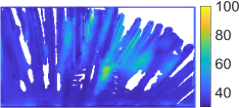}
\\
		\includegraphics[width=0.122\linewidth,keepaspectratio]{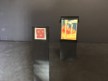} &
		\includegraphics[width=0.168\linewidth,keepaspectratio]{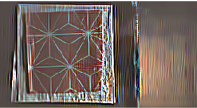} &
		\includegraphics[width=0.168\linewidth,keepaspectratio]{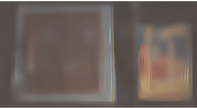} &
		\includegraphics[width=0.168\linewidth,keepaspectratio]{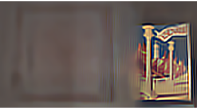} & 
		\includegraphics[width=0.168\linewidth,keepaspectratio]{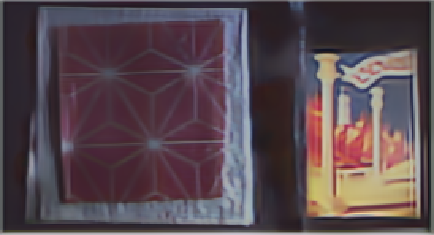} &
		\includegraphics[width=0.205\linewidth,keepaspectratio]{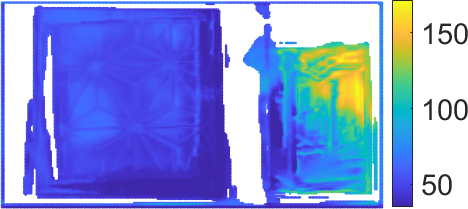}
\\
		\includegraphics[width=0.122\linewidth,keepaspectratio]{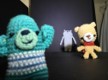} &
		\includegraphics[width=0.168\linewidth,keepaspectratio]{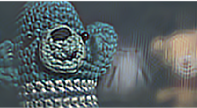} &
		\includegraphics[width=0.168\linewidth,keepaspectratio]{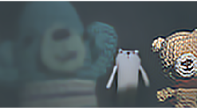} &
		\includegraphics[width=0.168\linewidth,keepaspectratio]{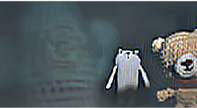} & 
		\includegraphics[width=0.168\linewidth,keepaspectratio]{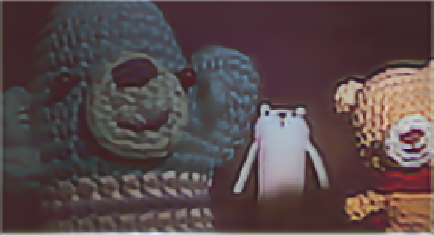} &
		\includegraphics[width=0.205\linewidth,keepaspectratio]{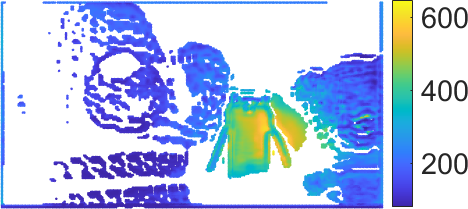}
	\end{tabular}
	\caption{Reconstruction of depth planes for different scenes using our proposed fast recovery algorithm with learned masks. Objects outside the recovered plane almost disappear. We also show all-in-focus image and depth map (in mm) created after passing the estimated multi-plane images through the trained U-net based refinement network, as discussed in Sec.~\ref{sec:denoiser_description}. Depth values of pixels with low intensity (e.g., mesh in the first row) are usually unrealiable and therefore removed. Results of additional depth planes and  scenes are available in the supplementary material. 
}
	\label{fig:results_learned}
\end{figure*}

\subsection{Camera Prototype}
To evaluate the performance of our proposed algorithms and mask patterns, we built a camera prototype (as shown in Figure~\ref{fig:overview} and the supplementary material). 
The camera consists of an image sensor and an LCoS display that acts as the programmable amplitude mask.
Our LCoS is HOLOEYE LC2012 transmissive spatial light modulator;  it has a pixel pitch of $36\mu$m and fill factor 58\%; and it is sandwiched between a pair of cross polarizers. The distance between LCoS and the sensor is 10.51mm.
We use a Sony IMX183 sensor on a board level Blackfly S, which has $5472\times 3648$ pixels with  $2.4\mu$m\ pitch and sensor area of 1.31cm $\times$ 0.88cm. In our experiments, we used $228\times342$ sensor measurements by binning $16\times 16$ adjacent pixels, which gives an effective sensor pixel of size $38.4\mu$m. 
The reconstructed image planes have $148\times274$ pixels because we crop the remaining pixels. 

We capture measurements for multiple scenes with varying depths in the range from 30mm to 600mm.
We calibrate the PSF of the camera at different depths. 
For each scene, we capture eight measurements using eight mask patterns with +1/-1 entries. We capture measurements for positive and negative parts of the masks separately and subtract the measurements for the negative part from the measurements for the positive part. Each mask has $63\times63$ pixels. We test the algorithms on four types of masks in our experiments: 
Shifted MLS masks: shifted versions of a single separable MLS mask \cite{hua2020sweep}. The shifting distance is from 0 to 48 LCoS pixels. 
MLS masks are separable MLS masks generated from different random seeds.
Random masks are non-separable random $\pm 1$ masks.
Learned masks are the he optimized masks generated from data-driven method discussed in Sec.~\ref{sec:mask_optimization}. To present the images with the same dynamic range, we applied the same brightness and color contrast adjustment to all the images.

\begin{figure}[htb]
	\setlength\tabcolsep{1pt}
	\centering
	\footnotesize
	\begin{tabular}{c|ccc}
			camera view &
		near plane &
		middle plane &
		far plane \\
		
		\includegraphics[width=0.19\linewidth,keepaspectratio]{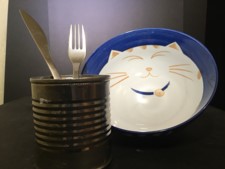} &
		\includegraphics[width=0.27\linewidth,keepaspectratio]{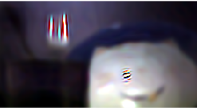} &
		\includegraphics[width=0.27\linewidth,keepaspectratio]{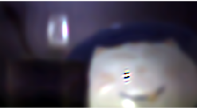} &
		\includegraphics[width=0.27\linewidth,keepaspectratio]{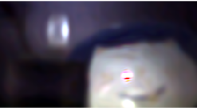}
		
	\end{tabular}
	\caption{Reconstructions of a scene with specular reflections that our method fails to recover. }
	\label{fig:results_learned_failure}
\end{figure}

\subsection{Reconstruction using Learned Masks}
We first present some example scenes and the reconstructed image planes using our proposed method. 
For each scene, we capture sensor measurements using eight learned mask patterns. Then we recover eight depth planes using the proposed algorithm in Sec.~\ref{sec:reconstruction_algorithm}. 
We sampled the depth planes uniformly in $[1-\frac{d}{z_\text{min}}, 1-\frac{d}{z_\text{max}}]$, where we input rough estimate of $z_\text{min},z_\text{max}$ for each scene.
Figure~\ref{fig:results_learned} presents three of the estimated planes for each scene. 
The first scene has a net mesh nearly 30mm from the camera and a rubber duck  nearly 100mm behind the net. The near-depth plane image is focused on the net and duck is blurred, while the far-depth plane is focused on the rubber duck and the net has almost disappeared. 
The third scene consists of 2 cards, the red card (on left) is placed at 35mm and the yellow card (on right) is placed at 200mm.  The near plane has the left card in focus and the far plane has the right card in focus. 
{The last scene has 3 toys, the blue toy is placed at 80mm, yellow toy at 200mm, and white toy at 600mm. We observe that the blue, yellow, and white toys appear sharpest in near, middle, and far planes, respectively. In lensless imaging systems, depth estimates of farther and darker objects are not as accurate as that of the closer and brighter objects; therefore, we removed the depth values of low-intensity pixels from the displayed depth maps.}

\renewcommand{\figwidth}{0.3\linewidth}
\begin{figure}[t]
	\setlength\tabcolsep{1pt}
	\centering
    \footnotesize
	\begin{tabular}{ccc}
			27.3mm &
		42.9mm &
		100.0mm \\
		\rotatebox{90}{\parbox{1.4cm}{\centering  SweepCam}}
		\includegraphics[width=\figwidth,keepaspectratio]{figures/net_d1_sweep.png} &
		\includegraphics[width=\figwidth,keepaspectratio]{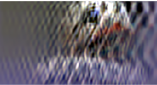} &
		\includegraphics[width=\figwidth,keepaspectratio]{figures/net_d3_sweep.png}
		\\
		\rotatebox{90}{\parbox{1.4cm}{\centering  \scriptsize Shifted MLS}}
		\includegraphics[width=\figwidth,keepaspectratio]{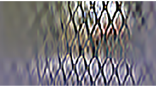} &
		\includegraphics[width=\figwidth,keepaspectratio]{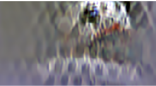} &
		\includegraphics[width=\figwidth,keepaspectratio]{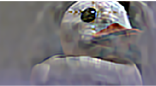}
		\\
		\rotatebox{90}{\parbox{1.4cm}{\centering  MLS}}
		\includegraphics[width=\figwidth,keepaspectratio]{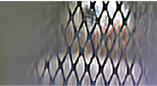} &
		\includegraphics[width=\figwidth,keepaspectratio]{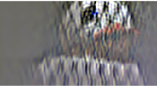} &
		\includegraphics[width=\figwidth,keepaspectratio]{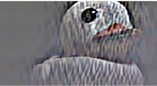}
		\\
		\rotatebox{90}{\parbox{1.4cm}{\centering  Random}}
		\includegraphics[width=\figwidth,keepaspectratio]{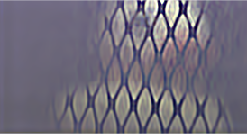} &
		\includegraphics[width=\figwidth,keepaspectratio]{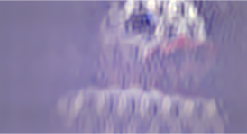} &
		\includegraphics[width=\figwidth,keepaspectratio]{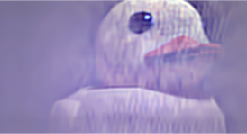}
		\\
		\rotatebox{90}{\parbox{1.4cm}{\centering  Learned}}
		\includegraphics[width=\figwidth,keepaspectratio]{figures/net_d1_learned.png} &
		\includegraphics[width=\figwidth,keepaspectratio]{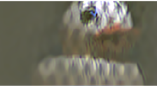} &
		\includegraphics[width=\figwidth,keepaspectratio]{figures/net_d3_learned.png}
		\\
	\end{tabular}
	\caption{Reconstructions of depth planes for recovered depth planes using SweepCam \cite{hua2020sweep} and our proposed methods using shifted MLS masks, MLS masks, random masks, and learned masks. We observe that the learned masks outperform all the other masks. }
	\label{fig:results_masks}
\end{figure}

The scene shown in Figure~\ref{fig:results_learned_failure} consists of a fork and a plate and demonstrates the limitation of our imaging system. Both of these objects have non-Lambertian surfaces and include specular reflections, which violates our convolution assumption. For this reason, the reconstructed images show artifacts.

\renewcommand{\figwidth}{0.14\linewidth}
\newcommand{\figwidthB}{0.17\linewidth}

\begin{figure*}[thb]
	\setlength\tabcolsep{1pt}
	\centering
	\footnotesize
	\begin{tabular}{cccccc}
			all-in-focus image &
		depth map &
		all-in-focus image & 
		depth map &
		all-in-focus image & 
		depth map\\
		\rotatebox{90}{\parbox{1.5cm}{\centering  Depth pursuit}}
		\includegraphics[width=\figwidth,keepaspectratio]{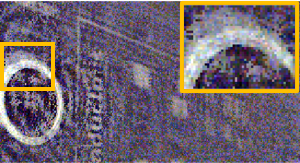} &
		\includegraphics[width=\figwidthB,keepaspectratio]{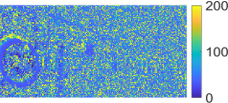} &
		\includegraphics[width=\figwidth,keepaspectratio]{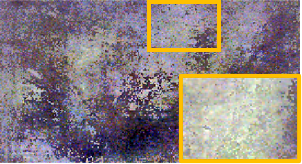}
		& 
		\includegraphics[width=\figwidthB,keepaspectratio]{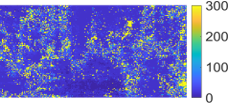} &
		\includegraphics[width=\figwidth,keepaspectratio]{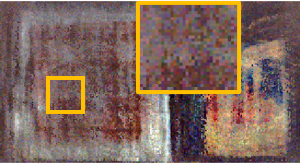}
		& 
		\includegraphics[width=\figwidthB,keepaspectratio]{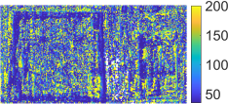}
		\\
		\rotatebox{90}{\parbox{1.5cm}{\centering  SweepCam}}
		\includegraphics[width=\figwidth,keepaspectratio]{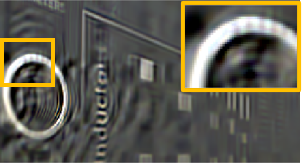} &
		\includegraphics[width=\figwidthB,keepaspectratio]{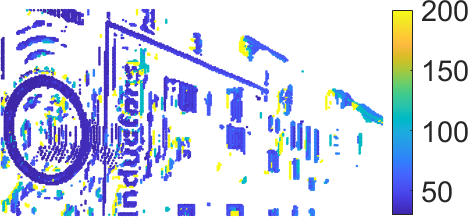} &
		\includegraphics[width=\figwidth,keepaspectratio]{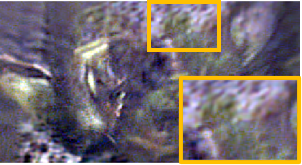}
		& 
		\includegraphics[width=\figwidthB,keepaspectratio]{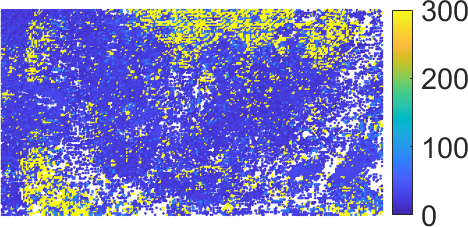}&
		\includegraphics[width=\figwidth,keepaspectratio]{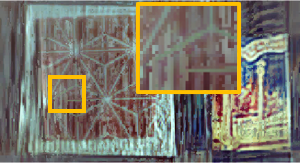}
		& 
		\includegraphics[width=\figwidthB,keepaspectratio]{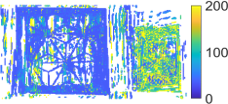}
		\\
		\rotatebox{90}{\parbox{1.5cm}{\centering  Ours}}
		\includegraphics[width=\figwidth,keepaspectratio]{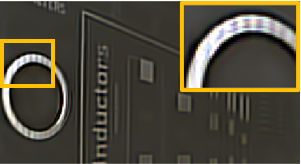} &
		\includegraphics[width=\figwidthB,keepaspectratio]{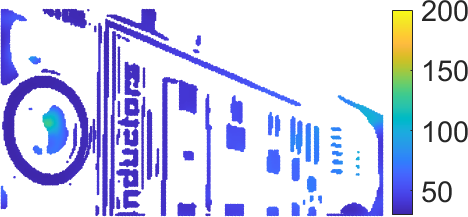} &
		\includegraphics[width=\figwidth,keepaspectratio]{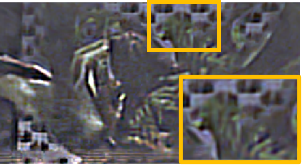}		& 
		\includegraphics[width=\figwidthB,keepaspectratio]{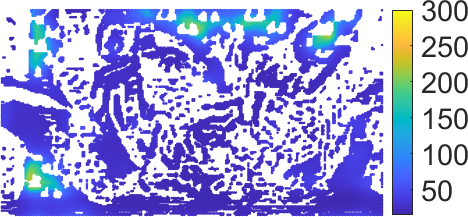} &
		\includegraphics[width=\figwidth,keepaspectratio]{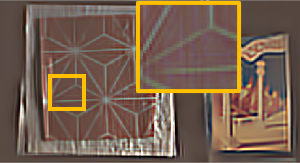}
		& 
		\includegraphics[width=\figwidthB,keepaspectratio]{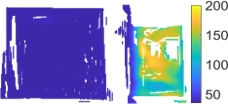}
	\end{tabular}
	\caption{Comparison of the depth pursuit algorithm \cite{asif2017lensless3D}, SweepCam  \cite{hua2020sweep}, and our proposed method. The details in the results from our proposed method are cleaner and sharper than the results from other methods. }
	\label{fig:results_algorithms}
\end{figure*}

\subsection{Comparison of Mask Patterns}
We show the comparison between different mask patterns and the learned masks. We present the reconstruction results in Figure~\ref{fig:results_masks}. 
We compare SweepCam method with shifted MLS masks \cite{hua2020sweep} and our method using shifted MLS masks, MLS masks, random masks, and learned masks. Figure~\ref{fig:results_masks} shows that SweepCam fails to recover depth planes accurately because we only used eight mask patterns. The reconstruction from learned masks using our method provides the best quality among all the mask types; far-depth plane separates the net and duck clearly. Compared to the learned masks, all the other mask patterns carry artifacts in their reconstructions (the shifted MLS and MLS masks give poor reconstruction for image at 100mm and random mask reconstruction exhibit haze artifacts). Additional experiments are available in the supplementary material.

\subsection{Comparison with Existing Methods}
We compare the performance of our proposed method with the greedy depth pursuit algorithm in \cite{asif2017lensless3D} and the depth sweep method in \cite{hua2020sweep}. 
% We capture measurements for multiple scenes within the depth range from 30mm to 600mm. 
In our experiments, the depth pursuit algorithm uses a single MLS mask, SweepCam method uses 8 shifted MLS masks, and our proposed method uses 8 learned masks.
We present the estimated all-in-focus images and  depth maps from the three methods in Figure~\ref{fig:results_algorithms}, which shows that our  method provides a better all-in-focus image and depth map compared to the other methods. 

\subsection{Refinement and Post-Processing}
\begin{figure}[ttt]
    \centering
	\setlength\tabcolsep{1pt}
    \footnotesize
    \begin{tabular}{cccc}
    \multicolumn{2}{c}{Local contrast} & \multicolumn{2}{c}{U-Net} \\ 
        all-in-focus image      &   depth map       &   all-in-focus image        &     depth map     \\
          \includegraphics[width=0.24\linewidth,keepaspectratio]{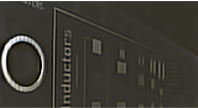}    &    
          \includegraphics[width=0.24\linewidth,keepaspectratio]{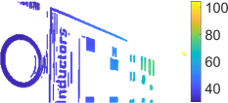}   &
          \includegraphics[width=0.24\linewidth,keepaspectratio]{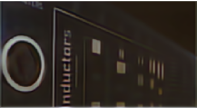}    &    
          \includegraphics[width=0.24\linewidth,keepaspectratio]{figures/board_depth_bd.png}   \\
          \includegraphics[width=0.24\linewidth,keepaspectratio]{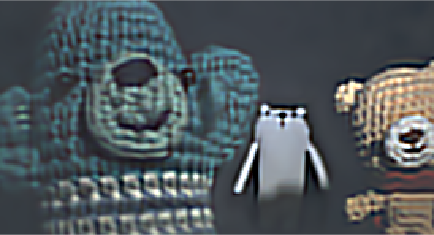}   &     
          \includegraphics[width=0.24\linewidth,keepaspectratio]{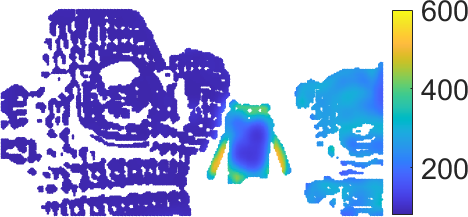} &
          \includegraphics[width=0.24\linewidth,keepaspectratio]{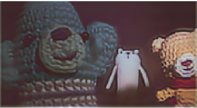}   &     
          \includegraphics[width=0.24\linewidth,keepaspectratio]{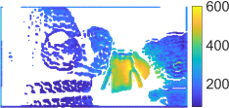} \\
    \end{tabular}
    \caption{The all-in-focus images and depth maps generated by the local contrast-based method and the trained U-Net. Local contrast method uses 47.85 seconds and U-Net uses 0.0043 seconds on average.
    }
    \label{fig:results_denoisers}
\end{figure}

We show a comparison between two post-processing approaches for converting estimated image planes to all-in-focus image and continuous-valued depth map in Figure~\ref{fig:results_denoisers}. In the first approach, we apply a BM3D denoiser \cite{dabov2007bm3d} on every estimated plane; then we perform a local contrast analysis to select the depth for every pixel that has maximum local contrast. In the second approach, we use a trained U-Net \cite{olaf2015unet} following the approach outlined in Sec.~\ref{sec:denoiser_description}.  Figure~\ref{fig:results_denoisers} shows that the U-Net results have less artifacts but all-in-focus image looks blurry. The results from local contrast-based method have more artifacts but images look sharp.

\subsection{Computational Complexity and Time}
To reconstruct an $M_u\times M_v\times D$ volume with $K$ mask patterns, we solve $M = M_uM_v$ least-squares problems, each of size $K\times D$ as in \eqref{eq:LS_parallel_sol}. The computational complexity of every least-squares problem is $\mathcal{O}(D^3+KD^2)$ \cite{golub1996matrix}. Since all the least-square problems can be solved independently of each other, we can solve them in parallel to accelerate the algorithm. 
In practice, the average running time to reconstruct eight depth planes with a single mask using the greedy depth pursuit algorithm \cite{asif2017lensless3D} is 138.16 seconds. The average running time to reconstruct eight depth planes with eight programmable masks are 0.82 seconds for SweepCam\cite{hua2020sweep} and 0.33 seconds for our method. In the post-processing step, which converts image planes to all-in-focus image and depth map, BM3D followed by local contrast method requires an average of 47.85 seconds, while the trained U-Net requires 0.0043 seconds.

\section{Conclusion}
In this paper, we present a new framework to recover 3D scenes using a lensless camera with a programmable mask. Our proposed method can recover multiple depth planes in the 3D scene using a computationally efficient algorithm that solves multiple small linear systems in parallel in the frequency domain. To further improve the quality of 3D scene recovery, we optimized the mask patterns and trained a U-Net that converts estimated image planes to all-in-focus image and continuous-valued depth map. Our experimental results demonstrate that the proposed method can reliably recover dense 3D scenes with a small number of sensor measurements and outperform existing methods. The reconstruction quality of our proposed method, like other lensless imaging systems, drops for scenes with specular reflections and large occlusions. 

\noindent Supplementary material is available at the following link: \href{https://github.com/CSIPlab/Programmable3Dcam.git}{https://github.com/CSIPlab/Programmable3Dcam.git}

\vspace{2pt}
\noindent \textbf{Acknowledgments.} This work was partially supported by the NSF grants CCF-2046293, CCF-1652569, and IIS-1730147 and the ONR grant N00014-19-1-2264.

{\small
\bibliographystyle{ieee_fullname}
\bibliography{egbib}

\begin{thebibliography}{10}\itemsep=-1pt

\bibitem{adams2017rice_depth}
Jesse~K. Adams, Vivek Boominathan, Benjamin~W. Avants, Daniel~G. Vercosa, Fan
  Ye, Richard~G. Baraniuk, Jacob~T. Robinson, and Ashok Veeraraghavan.
\newblock Single-frame 3d fluorescence microscopy with ultraminiature lensless
  flatscope.
\newblock {\em Science Advances}, 3(12), 2017.

\bibitem{antipa2018diffusercam}
Nick Antipa, Grace Kuo, Reinhard Heckel, Ben Mildenhall, Emrah Bostan, Ren Ng,
  and Laura Waller.
\newblock Diffusercam: lensless single-exposure 3d imaging.
\newblock {\em Optica}, 5(1):1--9, Jan 2018.

\bibitem{antipa2019still}
Nick Antipa, Patrick Oare, Emrah Bostan, Ren Ng, and Laura Waller.
\newblock Video from stills: Lensless imaging with rolling shutter.
\newblock In {\em IEEE International Conference on Computational Photography
  (ICCP)}, pages 1--8, Los Alamitos, CA, USA, may 2019. IEEE Computer Society.

\bibitem{asif2017lensless3D}
M.~Salman Asif.
\newblock Lensless 3d imaging using mask-based cameras.
\newblock In {\em IEEE International Conference on Acoustics, Speech and Signal
  Processing (ICASSP)}, pages 6498--6502. IEEE, April 2018.

\bibitem{asif2017flatcam}
M.~Salman Asif, Ali Ayremlou, Aswin~C. Sankaranarayanan, Ashok Veeraraghavan,
  and Richard Baraniuk.
\newblock Flatcam: Thin, lensless cameras using coded aperture and computation.
\newblock {\em IEEE Transactions on Computational Imaging}, 3(3):384--397, Sept
  2017.

\bibitem{boominathan2016lensless}
Vivek Boominathan, Jesse~K. Adams, M.~Salman Asif, Benjamin~W. Avants, Jacob~T.
  Robinson, Richard~G. Baraniuk, Aswin~C. Sankaranarayanan, and Ashok
  Veeraraghavan.
\newblock Lensless imaging: A computational renaissance.
\newblock {\em IEEE Signal Processing Magazine}, 33(5):23--35, 2016.

\bibitem{boominathan2020phlatcam}
Vivek Boominathan, Jesse~K. Adams, Jacob~T. Robinson, and Ashok Veeraraghavan.
\newblock Phlatcam: Designed phase-mask based thin lensless camera.
\newblock {\em IEEE Transactions on Pattern Analysis and Machine Intelligence},
  42(7):1618--1629, 2020.

\bibitem{busboom1998ura}
Axel Busboom, Harald Elders-Boll, and Hans~D. Schotten.
\newblock Uniformly redundant arrays.
\newblock {\em Experimental Astronomy}, 8(2):97--123, Jun 1998.

\bibitem{cannon1980coded_aperture}
Thomas~M. Cannon and Edward~E. Fenimore.
\newblock {Coded Aperture Imaging: Many Holes Make Light Work}.
\newblock {\em Optical Engineering}, 19:283, June 1980.

\bibitem{chang2019deep}
Julie Chang and Gordon Wetzstein.
\newblock Deep optics for monocular depth estimation and 3d object detection.
\newblock In {\em IEEE/CVF International Conference on Computer Vision (ICCV)},
  pages 10193--10202, 2019.

\bibitem{dabov2007bm3d}
Kostadin Dabov, Alessandro Foi, Vladimir Katkovnik, and Karen Egiazarian.
\newblock Image denoising by sparse 3-d transform-domain collaborative
  filtering.
\newblock {\em IEEE Transactions on Image Processing}, 16(8):2080--2095, 2007.

\bibitem{fenimore1978ura}
Edward~E. Fenimore and Thomas~M. Cannon.
\newblock Coded aperture imaging with uniformly redundant arrays.
\newblock {\em Applied Optics}, 17(3):337--347, Feb 1978.

\bibitem{galatsanos1991least}
Nikolas~P. Galatsanos, Aggelos~K. Katsaggelos, Roland~T. Chin, and Allen~D
  Hillery.
\newblock Least squares restoration of multichannel images.
\newblock {\em IEEE Transactions on Signal Processing}, 39(10):2222--2236,
  1991.

\bibitem{golomb2017shift}
Solomon~W Golomb.
\newblock {\em Shift register sequences: secure and limited-access code
  generators, efficiency code generators, prescribed property generators,
  mathematical models}.
\newblock World Scientific, 2017.

\bibitem{golub1996matrix}
Gene~H. Golub and Charles~F. Van~Loan.
\newblock {\em Matrix Computations}.
\newblock The Johns Hopkins University Press, third edition, 1996.

\bibitem{hua2020sweep}
Yi Hua, Shigeki Nakamura, M.~Salman Asif, and Aswin~C. Sankaranarayanan.
\newblock Sweepcam — depth-aware lensless imaging using programmable masks.
\newblock {\em IEEE Transactions on Pattern Analysis and Machine Intelligence},
  42(7):1606--1617, 2020.

\bibitem{khan2019iccv}
Salman~S. Khan, V.~R. Adarsh, Vivek Boominathan, Jasper Tan, Ashok
  Veeraraghavan, and Kaushik Mitra.
\newblock Towards photorealistic reconstruction of highly multiplexed lensless
  images.
\newblock In {\em IEEE/CVF International Conference on Computer Vision (ICCV)},
  pages 7859--7868, October 2019.

\bibitem{kingma2014adam}
Diederik~P. Kingma and Jimmy Ba.
\newblock Adam: {A} method for stochastic optimization.
\newblock In {\em International Conference on Learning Representations (ICLR)},
  2015.

\bibitem{metzler2020hdr}
Christopher~A. Metzler, Hayato Ikoma, Yifan Peng, and Gordon Wetzstein.
\newblock Deep optics for single-shot high-dynamic-range imaging.
\newblock In {\em IEEE/CVF Conference on Computer Vision and Pattern
  Recognition (CVPR)}, 2020.

\bibitem{monakhova2020spectral}
Kristina Monakhova, Kyrollos Yanny, Neerja Aggarwal, and Laura Waller.
\newblock Spectral diffusercam: Lensless snapshot hyperspectral imaging with a
  spectral filter array.
\newblock {\em Optica}, 7(10):1298--1307, 2020.

\bibitem{monakhova2019learning}
Kristina Monakhova, Joshua Yurtsever, Grace Kuo, Nick Antipa, Kyrollos Yanny,
  and Laura Waller.
\newblock Learned reconstructions for practical mask-based lensless imaging.
\newblock {\em Opt. Express}, 27(20):28075--28090, Sep 2019.

\bibitem{silberman2012NYU_depth}
Pushmeet~Kohli Nathan~Silberman, Derek~Hoiem and Rob Fergus.
\newblock Indoor segmentation and support inference from rgbd images.
\newblock In {\em European Conference on Computer Vision (ECCV)}, 2012.

\bibitem{canh2019iccv}
Thuong Nguyen~Canh and Hajime Nagahara.
\newblock Deep compressive sensing for visual privacy protection in flatcam
  imaging.
\newblock In {\em IEEE/CVF International Conference on Computer Vision (ICCV)
  Workshops}, Oct 2019.

\bibitem{sony2020infrared}
Ilya Reshetouski, Hideki Oyaizu, Kenichiro Nakamura, Ryuta Satoh, Suguru
  Ushiki, Ryuichi Tadano, Atsushi Ito, and Jun Murayama.
\newblock Lensless imaging with focusing sparse ura masks in long-wave infrared
  and its application for human detection.
\newblock In {\em European Conference on Computer Vision (ECCV)}, pages
  237--253, 2020.

\bibitem{olaf2015unet}
Olaf Ronneberger, Philipp Fischer, and Thomas Brox.
\newblock U-net: Convolutional networks for biomedical image segmentation.
\newblock In {\em International Conference on Medical Image Computing and
  Computer Assisted Intervention (MICCAI)}, pages 234--241, 2015.

\bibitem{scharstein2001stereo}
Daniel Scharstein, Richard Szeliski, and Ramin Zabih.
\newblock A taxonomy and evaluation of dense two-frame stereo correspondence
  algorithms.
\newblock In {\em IEEE Workshop on Stereo and Multi-Baseline Vision}, pages
  131--140, Dec 2001.

\bibitem{sitzmann2018deep}
Vincent Sitzmann, Steven Diamond, Yifan Peng, Xiong Dun, Stephen Boyd, Wolfgang
  Heidrich, Felix Heide, and Gordon Wetzstein.
\newblock End-to-end optimization of optics and image processing for achromatic
  extended depth of field and super-resolution imaging.
\newblock {\em ACM Transactions on Graphics (SIGGRAPH)}, 2018.

\bibitem{sun2020learning}
Qilin Sun, Ethan Tseng, Qiang Fu, Wolfgang Heidrich, and Felix Heide.
\newblock Learning rank-1 diffractive optics for single-shot high dynamic range
  imaging.
\newblock In {\em IEEE/CVF Conference on Computer Vision and Pattern
  Recognition (CVPR)}, June 2020.

\bibitem{tan2019face}
Jasper Tan, Li Niu, Jesse~K. Adams, Vivek Boominathan, Jacob~T. Robinson,
  Richard~G. Baraniuk, and Ashok Veeraraghavan.
\newblock Face detection and verification using lensless cameras.
\newblock {\em IEEE Transactions on Computational Imaging}, 5(2):180--194, June
  2019.

\bibitem{gordon2020deep}
Gordon Wetzstein, Aydogan Ozcan, Sylvain Gigan, Shanhui Fan, Dirk Englund,
  Marin Soljačić, Cornelia Denz, David A~B Miller, and Demetri Psaltis.
\newblock Inference in artificial intelligence with deep optics and photonics.
\newblock {\em Nature}, 588(7836):39—47, December 2020.

\bibitem{wu2019phasecam}
Yicheng Wu, Vivek Boominathan, Huaijin Chen, Aswin~C. Sankaranarayanan, and
  Ashok Veeraraghavan.
\newblock Phasecam3d — learning phase masks for passive single view depth
  estimation.
\newblock In {\em IEEE International Conference on Computational Photography
  (ICCP)}, pages 1--12, May 2019.

\bibitem{yamaguchi2019lensless}
Keita Yamaguchi, Yusuke Nakamura, Kazuyuki Tajima, Toshiki Ishii, Koji
  Yamasaki, and Takeshi Shimano.
\newblock Lensless 3d sensing technology with fresnel zone aperture based
  light-field imaging.
\newblock In {\em Industrial Optical Devices and Systems}, volume 11125, page
  111250F. International Society for Optics and Photonics, 2019.

\bibitem{zheng2019depth}
Yucheng Zheng and M. {Salman Asif}.
\newblock Joint image and depth estimation with mask-based lensless cameras.
\newblock {\em IEEE Transactions on Computational Imaging}, 6:1167--1178, 2020.

\end{thebibliography}
}
% \balance 

\end{document}